%
%
%

\ifx\mnmacrosloaded\undefined \input mn\fi

\def\der#1#2{{\partial#1\over\partial#2}}

\def\derss#1#2{{\partial^2#1\over\partial#2^2}}
\def\dert#1#2{{d#1\over d#2}}
\def\dertt#1#2{{d^2#1\over d#2^2}}

\def\kes{\kappa_{es}}
\def\mincir{\raise -2.truept\hbox{\rlap{\hbox{$\sim$}}\raise5.truept
\hbox{$<$}\ }}
\def\magcir{\raise -4.truept\hbox{\rlap{\hbox{$\sim$}}\raise5.truept
\hbox{$>$}\ }}
\def\rho{\varrho}

\def\ref{\bibitem}


\pageoffset{-2.5pc}{0pc}

%
  
%
%
%



\onecolumn        
\pagerange{0--0}    
\pubyear{0000}
\volume{000}

\begintopmatter  

\title{Dynamical Comptonization in spherical flows: black hole
accretion and stellar winds}

\author{R.~Turolla$^1$, S.~Zane$^2$, L.~Zampieri$^2$ and L.~Nobili$^1$}
\affiliation{$^1$ Department of Physics, University of Padova, Via Marzolo 8,
35131 Padova, Italy}
\smallskip
\affiliation{$^2$ International School for Advanced Studies (SISSA),
Via Beirut 2--4, 34013 Trieste, Italy}

\shortauthor{R. Turolla, S. Zane, L. Zampieri and L. Nobili}

\shorttitle{Comptonization in spherical flows}


\acceptedline{Accepted ... Received ... ; in original form ...}

\abstract{The transport of photons in steady, spherical, scattering flows 
is investigated. The moment equations are solved analytically for accretion 
onto a Schwarzschild black hole, taking into full account relativistic 
effects. We show that the emergent radiation spectrum is a power law at high 
frequencies with a spectral index smaller (harder spectrum) than in the 
non--relativistic case. Radiative transfer in an expanding envelope is
also analyzed. We find that adiabatic expansion produces a drift of 
injected monochromatic photons towards lower frequencies and the formation of
a power--law, low--energy tail with spectral index $-3$.
}

\keywords {accretion, accretion discs -- radiation mechanisms -- radiative 
transfer.}

\maketitle  

\section{Introduction}

It was realized long ago (Cowsik \& Lee 1982 and references therein) that the 
divergence of the velocity field in astrophysical flows can provide a very
efficient mechanism to transfer energy from the fluid to particles (photons,
neutrinos, cosmic rays) diffusing through the medium, even in the absence of
shocks. In the case of photons undergoing multiple scatterings off cold  
electrons, this effect is germane to thermal Comptonization, with the 
flow velocity $v$ playing the role of the 
thermal velocity, and is sometimes 
referred to as dynamical Comptonization. In a series of papers
Blandford \& Payne (1981a,b; Payne \& Blandford 1981, PB in the following) 
were the first to emphasize 
the importance of repeated scatterings in a steady, spherical 
flow of depth $\tau\gg 1$. They have shown that monochromatic photons injected
in a region where $\tau v/c\sim 1$ always gain energy, because of adiabatic
compression, and emerge with a broad
distribution which exhibits a distinctive power--law, high--energy tail.
Under the assumption that $v\propto r^\beta$, the spectral index depends only
on $\beta$. 

Cowsik \& Lee (1982), and later Schneider \& Bogdan (1989), stressed that 
Blandford
\& Payne diffusion equation for the photon occupation number could be
regarded as a particular case of the standard cosmic--ray transport 
equation in which the diffusion coefficient, $\kappa \propto r^\alpha
\nu^\gamma$, does not depend on the photon energy ($\gamma=0$) and $\alpha
-\beta = 2$. Starting from this result, Schneider \& Bogdan were
able to generalize PB analysis to include the transition between Thomson
($\gamma =0$) and Klein--Nishina ($\gamma =1$) scattering cross--sections.
The competitive role of dynamical and thermal Comptonization in accretion 
flows with a non--zero electron temperature was studied by Colpi (1988).
More recently, Mastichiadis \& Kylafis (1992) investigated the effects of 
dynamical Comptonization in near--critical accretion onto a neutron star. Their
approach is very similar to PB, but the presence of a perfectly reflecting
inner boundary (either the NS surface or the magnetosphere) was taken into
account. They have shown that, in this case, the emergent spectrum is much 
harder than in PB and that the spectral index depends both on $\beta$ and
on the depth at the inner boundary.

Even when dealing with black hole accretion, all previous analyses neglected
both special and general relativity. The presence of an event horizon was not 
considered and only terms up to first order in $v/c$ were retained.
In this paper we extend PB calculations to account for relativistic effects
on radiative transfer which arise when the flow velocity approaches the
speed of light in the vicinity of the hole horizon; as in PB, we assume that
scattering is elastic in the electron rest frame. 
The motivation for
this work is twofold, much like in Mastichiadis \& Kylafis. First: to 
investigate the properties of the emergent spectrum in a more realistic 
accretion scenario, in which
the optical depth near the horizon is not so large to prevent photons from 
escaping. Second: to check by means of an analytical calculation the results 
obtained with a numerical code recently developed for solving the complete 
transfer problem in spherical flows (Zane {\it et al.\/} 1996). Computed
spectra show, in fact, a power--law, high--energy tail but the spectral index
depends on
the optical depth at the horizon, even for fixed $\beta=-1/2$, and it is always
smaller than $2$ (PB result). Here we show that advection/aberration effects
in the high--speed flow near the horizon, due to the finiteness of the depth
there, produce a power--law tail flatter with respect to PB and enable photons
to drift also towards energies lower than the injection energy. 

In addition, we present an analysis of dynamical Comptonization in an
expanding atmosphere, using the same assumptions of PB. We found that the
solution for the emergent flux shows specular features with respect to PB.
Adiabatic expansion produces a drift of injected monochromatic photons to 
lower energies and the formation of  low--energy, power--law tail.
In this case, however, the spectral index is independent on the 
velocity gradient and turns out to be always equal to $-3$.

\section{Radiative transfer in a converging flow}

In this and in the following sections we deal with the transfer of radiation
through a scattering, steady, spherical flow, characterized by a power--law 
velocity
profile $v \propto r^{\beta}$. Under these assumptions the rest--mass
conservation yields immediately a density profile 
$\rho \propto r^{-2-\beta}$
from which it follows that the electron--scattering optical depth is 
$\tau = \kes\rho r/(1+\beta) \propto r^{-1-\beta}$.
The parameter $\beta$ is positive for outflows and it has to be $\beta > -1$ 
for the optical depth to decrease with increasing $r$. Units in which 
$c=G=1$ are used unless explicitly stated.  

Blandford \& Payne (1981a, b) and PB restricted their analysis to 
converging flows with 
a non--relativistic bulk velocity and to conservative and isotropic scattering
in the electron rest frame.
Combining the first two moment equations, written in an inertial frame, they
found that, in diffusion approximation, the (angle--averaged)
photon occupation number $n$ obeys a Fokker--Planck equation. Defining 
$\tau^* = 3 \tau v $ and looking for separable solutions of the form 
$$
n\left ( \nu, \tau^* \right ) = f \left ( \tau^* \right )
{\tau^*}^{3 +\beta} \nu^{-\lambda} \, ,  
\eqno (1) $$
the resulting second order ordinary 
differential equation for $f$ reduces to a confluent 
hypergeometric equation
(here the sign of $\beta$ is the opposite with respect to PB, 
e.g. $\beta = -1/2$ for free--fall, according to the assumptions at the 
beginning of this section).
As discussed by PB, 
the solution corresponding to a constant radiative flux at infinity and 
to adiabatic compression of photons for $\tau \to \infty$
is expressed in terms of the Laguerre polynomial
$L_n^{(3+\beta)}(\tau^*)$. The above two conditions give rise to a discrete 
set of eigenvalues for 
the photon index $\lambda$ which are given by 
$$
\lambda_n = { 3 \left ( n + 3 +\beta \right ) \over \left ( 2 +\beta 
\right ) }\qquad\qquad (n = 0,1,2,\ldots )\eqno (2) $$

The general solution is written as the superposition of different modes.
Assuming that monochromatic photons with $\nu =\nu_0$ are injected at 
$\tau^* = \tau^*_0$, it is 
$$
n \propto {\tau^*}^{3+\beta}\sum_{n=0}^\infty
{ \Gamma(n+1) \over \Gamma (n+4+ \beta )} L_n^{(3 + \beta )} (\tau^*_0)   
L_n^{(3 + \beta )} (\tau^*)\left( { \nu \over \nu_0}\right)^{-\lambda_n}\, .
\eqno (3)
$$
In this case, bulk motion comptonization tends to create a power law, high 
energy tail. Defining the spectral index as
$$
\alpha = - \dert {\ln L \left ( \nu,0 \right )}{\ln \nu} \eqno (4) $$
where $L$ is the luminosity, PB found 
$$
\lim_{\nu \to \infty} \alpha = { 3 \over 2 +\beta }
\eqno (5) $$
showing that the spectral slope at high frequencies is 
dominated by the fundamental mode $n=0$. In particular, it is
$\alpha =  2$ for a free--falling gas. 

The same results can be recovered using the PSTF moment formalism introduced 
by Thorne (1981), as shown by Nobili, Turolla
\& Zampieri (1993). Here we outline the general method,
mainly to introduce same basic concepts that will be used
in the following sections. In particular, we consider 
the first two PSTF moment equations with only Thomson scattering included in 
the source term. In the frame comoving with the fluid they read 
$$
\der{w_1}{\ln r} + 2w_1   + { y' \over y } \left(
w_1 - \der{w_1}{\ln\nu} \right)
-v \left[\der{w_0}{\ln r} + 
\left( 1 - \beta \right)\der{w_2}{\ln\nu}
- (2+\beta ) \left ({1\over 3} \der{w_0}{\ln\nu} - w_0 \right ) 
\right] = 0  \eqno(6a)
$$

$$\eqalign {&
{1\over 3}\der{w_0}{\ln r} + 
\der{w_2}{\ln r }
+ 3 w_2  - {y' \over y} \der{w_2}{\ln\nu} +
 { y'\over y}\left(w_0 - {1\over 3}\der{w_0}{\ln\nu}\right) -
v\left[ - {3\over 5}(4+\beta )w_1 - 
{1\over 5} (2+ 3\beta )
\der{w_1}{\ln\nu} + \der{w_1}{\ln r} + 
\right. \cr
&
\left. \left(1 - \beta \right) 
\left( w_3+ \der{w_3}{\ln\nu}\right)\right] +
{{(1+\beta )\tau}\over y}w_1=0\, .\cr}\eqno(6b)$$
where a prime denotes the total derivative wrt $\ln r$, 
$ y = \sqrt{1 - r_g/r}/\sqrt {1 - v^2}$, $r_g$ is the gravitational radius, 
and $v$ is taken positive for inward motion.
As discussed by Turolla \& Nobili (1988, see also Thorne, Flammang \& 
\.Zytkow 1981), in diffusion approximation the hierarchy of 
the frequency--integrated PSTF moments $W_l$ is such that 
$$
\eqalign{
W_1 & \sim {W_0\over \tau} \cr
W_2 & \sim {W_0\over \tau}\left({1\over \tau} -v\right) \cr
W_3 & \sim {W_0 \over \tau^2} \left({1\over \tau} -v\right) \, . 
 \cr }\eqno(7)$$
The same hierarchy can be assumed to hold also for frequency--dependent 
moments in a scattering medium. 

PB result can be reproduced in the limit of large $\tau$ and small $v$, 
i.e. retaining in equations (6) 
only terms of order $w_0$, $v w_0$ and $w_0/ \tau$, and suppressing gravity,
which is equivalent to set $y=1$ and $y'=0$. In particular, 
under such hypothesis, all terms containing both $w_2$ and $w_3$ can be 
neglected and the moment equations become 
$$
\der{w_1}{\ln t} - v\der{w_0}{\ln t} - 2w_1 + v(2+\beta )\left(w_0
-{1\over 3}\der{w_0}{\ln\nu}\right)=0  \eqno(8a)
$$
$$
v\der{w_0}{\ln t} - tw_1=0\, , \eqno(8b)
$$
where $t = (1+\beta)\tau^*$.
Equations (8a) and (8b) can be combined together to yield a 
second order, partial differential equation for the radiative flux 
$$t\derss{w_1}{t} -\left(t+1-\beta\right )\der{w_1}{t}+\left(1-{{2\beta
\over t}}\right)w_1 - {{2+\beta}\over 3}\der{w_1}{\ln\nu}=0\, .\eqno(9)$$

Following PB, the solution of equation (9) can be found by
separation of variables. Writing $w_1 = t^p h_1(t)\nu^{-\alpha}$, it is easy
to show that for $p=2$ and for $p=-\beta$, equation (9) becomes 
a confluent hypergeometric equation for $h_1$. Actually the requirement of 
constant radiative flux at infinity is met only for $p=2$, and in this case
we get
$$
t\dertt{h_1}{t} +\left(3+\beta-t\right )\dert{h_1}{t}-\left(1-{{2+\beta}
\over 3}\alpha\right)h_1 = 0 \, . \eqno(10)
$$

As previously discussed, the physical solution for $w_1$ can be obtained as a
superposition of the Kummer functions $M(-n,3+\beta ,t)\propto 
L_n^{(2+\beta)}(t)$, for $n=0,1,\ldots$,
with corresponding eigenvalues
$$ \alpha_n =  {{3(n+1)}\over {2+\beta}} \, . \eqno(11) $$
In this case
$$ w_1 = t^2 \sum_{n=0}^{\infty} A_n L_n^{(2+\beta)}(t)\nu^{-\alpha_n} \, ,
\eqno(12)$$
where the $A_n$'s are constants to be fixed by the boundary conditions. 
At sufficiently large frequencies the spectrum is dominated by
the fundamental mode; in particular, for $\beta = -1/2$, it is again
$\alpha_0 = 2 \, .$

\section{Importance of relativistic effects}

Here we consider the effects of 
dynamical Comptonization in spherical accretion onto a non--rotating black 
hole, taking into full account both gravity and velocity terms in the 
moment equations. With reference to this particular problem, we can
safely assume that  
matter is free--falling, $v=(r/r_g)^\beta$ with $\beta = -1/2$. Note that
in spherical accretion onto black holes the radiative flux is never
going to influence the flow dynamics close to the horizon (see e.g. 
Gilden \& Wheeler 1980; Nobili, Turolla \& Zampieri 1991; Zampieri, Miller \& 
Turolla 1996). In this
case dynamics cancels, locally, gravity, so that 
$y=1$, $y'=0$. The moment equations look then ``non--relativistic'' in form, 
although $v$ can be arbitrarily close to unity. Corrections due to large
values of the flow velocity were not considered in previous works despite 
the fact that they are bounded to become important near the event horizon where 
$v\sim 1$. We note that the bulk of the emission in realistic accretion models
is expected to come precisely from the region close to $r_g$.
As in the non--relativistic analysis presented in the last
section, we consider the diffusion limit, truncating self--consistently
both equations (6) to terms of order $w_0/\tau$. 
The moments hierarchy, expressions (7), shows that all terms containing $w_3$ 
can be always neglected in equation (6b), since they are of order $w_0/\tau^2$. 
In the present case, however, all other terms must be retained. In fact, 
$v w_1\sim w_1\sim w_0/\tau$ when $v\sim 1$ and $w_2\sim w_0/\tau$,
at least for $v\sim 1\magcir 1/\tau$ (see again expressions [7]). 
This implies that $w_1$ and $w_2$ contribute to the
same extent to the anisotropy of the radiation field.
Note that under such conditions it is $W_2=4\pi(K-J/3)<0$, as already pointed 
out by Turolla \& Nobili (1988), so that in high--speed, diffusive flows $K$ 
may become less than $J/3$. Contrary to the
case discussed in section 2 where $w_2$ is negligible, now the system 
of the first two moment equations is not closed. However, up to terms of 
order $w_0/\tau$, the second moment equation
does not contain moments of order higher than $w_2$ and provides then 
the required closure equation
$$
v\der{w_2}{\ln t} - {4\over 15}\der{w_1}{\ln t}   -{15\over 14}vw_2-{4\over 15}
w_1 + {2\over 5}vw_0 + 
{3\over 14}v\der{w_2}{\ln\nu} - {2\over 15}v\der{w_0}
{\ln\nu} + {3\over 10}{t\over v}w_2 = 0\, . \eqno(13)
$$
The complete system (6a), (6b) and (13) is awkward and a solution can be 
obtained only numerically. It is possible, nevertheless, to find an
analytical solution if we consider the closure condition for $w_2$ which
follows from equation (13) with only terms of order $w_0$ retained
$$w_2 = {4\over 9}{v\over t}\left(\der{w_0}{\ln\nu}-3w_0\right)\, .
\eqno(14)$$
With this closure, $w_2$ is always negative provided that
$\partial w_0/\partial\ln\nu <0$. This implies that equation (14) is strictly 
valid only for $\tau v
\magcir 1$ (see expression [7]), that is to say below the trapping radius.
Introducing the new dependent variables $f_0 = vw_0$, $f_1 = w_1$ and
$f_2 = vw_2$, the moment equations become
$$
t\der{f_0}{t} + {1 \over 2} \der{f_0}{\log \nu} - 2 f_0 - t\der{f_1}{t}
+ 2 f_1 - { 3 \over 2} \der{f_2}{\log \nu} = 0 \eqno(15a)
$$
$$
{t \over 3} \der { f_0}{t} - { 1 \over 6 } f_0 - { t \over t_h} 
\left [ t\der { f_1}{t} + { 1\over 10} \der {f_1} {\log \nu}  + \left ( 
{t_h\over 3}  - { 9 \over 10} \right ) f_1  \right ] 
 + t\der{f_2}{t} - { 7 \over 2 } f_2 = 0 \eqno(15b)
$$
$$
f_2  - { 4  \over 9 t_h } \left( \der{f_0}{\log \nu} - 3 f_0\right) = 0\, ,
\eqno(15c) 
$$
where $t_h$ is the value of $t$ at the radius where $v=1$, i.e. at $r=r_g$ in
the case under examination.
We note that for $t_h \to \infty$ 
equations (15a,b) give exactly the low--velocity limit of PB 
(equations [8a,b]), irrespective of the value of $v$. This is because when
$t_h\to\infty$ the scattering depth itself near the horizon must be very 
large, 
so the radiation field there is very nearly isotropic. Departures from isotropy,
due both to the radiative flux $w_1\sim w_0/\tau$ and to the radiative shear
$w_2\sim (1/\tau - v)w_0/\tau$ become 
vanishingly small, no matter how large velocity is. Under such conditions
PB approach is still valid just because {\it both\/} $w_1$ and $w_2$ become
negligible in the moment equations, although they may be of the same order.

The system (15) can be solved looking again for separable 
solutions of the type $f_i = g_i(t) \nu^{-\alpha}$.
After some manipulation, it can be transformed into a pair of
decoupled, second order, ordinary  differential equation for $g_0(t)$ and
$g_1(t)$, having the same structure. In particular for $g_1(t)$ it is
$$
t^2 \left ( \beta t + \gamma \right ) \dertt{g_1}{t} + t 
\left ( \delta t + \epsilon \right ) \dert {g_1}{t} 
- \left ( \eta t + \lambda \right ) g_1 = 0\eqno (16)
$$
where 
$$\eqalign {
& \beta = 60 t_h\, , \cr
& \gamma = -(20/3) t_h \left [ 3 t_h - 4 \left ( 3 + \alpha \right ) \right ]
\, ,  \cr
& \delta = 20 t_h^2 - 18 \left ( 2 \alpha + 3 \right ) t_h - 40 
\alpha \left ( 3 + \alpha \right )\, ,  \cr
& \epsilon = 30 t_h \left [ t_h - 4 \left ( 3 + \alpha \right ) \right ]
\, , \cr
& \eta = 10 \left ( \alpha + 2 \right ) t_h^2 + \left ( 31 \alpha^2/3 
+ 7 \alpha - 54 \right )t_h - 4 \alpha \left (\alpha + 3 \right ) 
\left ( \alpha + 9 \right )\, ,  \cr
& \lambda = (20/3) t_h \left [ 3 t_h - 28 \left (3 + \alpha \right ) \right ]\,
. \cr}
$$
Equation (16) can be reduced to a hypergeometric equation upon the 
change of variables $g_1(t) = t^p h_1(z)$ and $ z = - (\beta/\gamma)t$, 
where $p$
is the solution of the quadratic equation $\gamma p^2 + (\epsilon - \gamma)p -
\lambda = 0$.
A direct, but tedious, calculation shows that 
$$\eqalign{ 
         & p_+ = 2 \cr
         & p_- = {7\over 2} -{9 t_h \over 3t_h -4(3 + \alpha)}\, ;\cr}
\eqno(18)
$$  
in the limit $t_h \to \infty$, $p_- = 1/2$
as in the case considered in the previous section,
and $p=p_+$ will be used in the following to meet the requirement of constant 
radiative flux at infinity. Equation (16) can be now written in 
the form
$$
z ( 1 - z) \dertt{h_1}{z} + \left [ { \epsilon \over \gamma} + 2 p - \left ( 
{ \delta \over \beta} + 2 p \right ) z \right ] \dert{h_1}{z} - \left [ 
p \left ( p -1 \right ) + p { \delta \over \beta } - { \eta \over \beta } 
\right ] h_1 = 0 \eqno (19)
$$
which is a hypergeometric equation. The general solution is expressed in terms
of the hypergeometric function $\null_2F_1( a, b, c; z)$ and the three 
parameters $a$, $b$, $c$
(see Abramowitz \& Stegun 1972, AS in the following, for notation) are given 
by the relations
$$\eqalign{ 
&  c = {\epsilon\over\gamma} + 2p\cr
&  a+b +1  = {\delta\over\beta} + 2p\cr
&  ab  = p ( p-1) + p {\delta\over\beta} - {\eta\over\beta}\, .\cr} 
$$
Solving for $a$, $b$ we obtain, after a considerable amount of algebra, 
$$
a = { 2 -\alpha  \over 2} - {2 \alpha \left ( \alpha + 3 \right ) \over 3 t_h}
\eqno(20a)
$$
$$
b = {t_h\over 3} + {11 - \alpha \over 10}\, . \eqno(20b)
$$  

It can be seen from equations (20) that, in the limit $t_h \to \infty$, $b$ 
diverges while $a$ stays finite; in the same limit 
the hypergeometric equation reduces to 
the confluent hypergeometric equation (see e.g. Sneddon 1956). As discussed 
in section 2, the relevant
solution in the non--relativistic case is given by Laguerre polynomials
and is recovered imposing $a = -n$, with $n=0,1,\ldots$. 
The solution of equation (21) which reduces to PB for $t_h\to\infty$ is found 
imposing again that $a$ is either zero or a negative integer (although other 
classes of solutions that do not match PB may exist).
In this case $h_1$ is still polynomial and takes the form
$$
h_1 (z) = \null_2F_1 \left ( -n , b, c; z \right ) 
= { n ! \over \left (c \right )_n } P_n^{\left (c-1,b-c-n \right )} 
\left ( 1 - 2z \right ) \, , \eqno (21)
$$
where $P_n^{(p,q)}(z)$ is the Jacobi polynomial and $(c)_n = 
\Gamma(c +n)/\Gamma(c)$ is the 
Pochhammer's symbol (see again AS). For $t_h\to\infty$, it is 
$c\sim 5/2$, $b\sim t_h/3$, $z\sim 3t/t_h=t/b$ and 
$$P_n^{(c-1,b-c-n)}\left(1-2{t\over b}\right)\to L_n^{(3/2)}(t)\, ,$$
so that, as expected, the solution of section 2 is recovered.

The discrete set of eigenvalues for the spectral index $\alpha_n$ follows
immediately from (20a) solving the quadratic equation $a = -n$. For each
$n$ both a positive, $\alpha^+_n$, and a negative, $\alpha^-_n$, mode is
present
$$
\alpha^{\pm}_n =  {- (12 + 3 t_h) \pm \sqrt{ \left ( 12 + 3 t_h \right )^2 + 
96(n+1)t_h }\over 8 } \, . \eqno (22)
$$
We checked that the eigenvalues of the equation for $g_0$ are again given by 
equation (22), in agreement with the starting hypothesis that $\alpha$ is the 
same for all moments.  

The general solution for the spectral flux is obtained as a linear
superposition of all modes
$$
w_1 = t^2 \left[\sum_{n=0}^\infty A^+_n(-1)^n{{(b)_n}\over{(c)_n}}
G_n(b-n,c,z)\left({\nu\over\nu_0}\right)^{-\alpha^+_n} + 
\sum_{n=0}^\infty A^-_n(-1)^n{{(b)_n}\over{(c)_n}}
G_n(b-n,c,z)\left({\nu\over\nu_0}\right)^{-\alpha^-_n}\right]\, , \eqno(23)
$$
where we have expressed $h_1$ in terms of the shifted Jacobi polynomials
$G_n$. We remind that $b$, $c$ and $z$ are all functions of $\alpha^{\pm}_n$,
although we dropped all indices to simplify the notation. The two sets of
constants $A^\pm_n$ are fixed imposing a boundary condition at the injection
frequency $\nu=\nu_0$. The only boundary condition compatible with
the assumption of a pure scattering flow for $t< t_h$ is that all photons are 
created in an infinitely thin shell at $t_*$. This is equivalent to ask that 
$w_1(t,\nu_0 )\propto\delta(t/t_* -1)$, as in PB.

At variance with the results discussed in section 2, now the series in equation
(23) can not be summed using the polynomial generating function because $b$, 
$c$ and $z$ depend on $n$. The coefficients $A^\pm_n$ are
solution of an upper triangular, infinite system of linear 
algebraic equations (see Appendix A). It can be easily shown that the
two series in equation (23) do not converge for any value of $\nu/\nu_0$.
In fact, the general term of the first series, which is of the type 
$f(n)(\nu_0/\nu)^{\alpha^+_n}$, can not be infinitesimal for arbitrarily
small frequencies unless the series truncates, which is not the case if
it must reproduce the $\delta$--function at $\nu=\nu_0$. On the other 
hand, the series is absolutely convergent for $\nu > \nu_0$,
provided that $|f(n)|$ is bounded. For $N\gg 1$, the series has 
a majorant $\propto\sum_{n=N}^\infty(\nu_0/\nu)^{\sqrt n}$ which
is convergent because $\int_N^\infty(\nu_0/\nu )^{\sqrt x}\, dx$ is finite 
for $\nu > \nu_0$. 
The same argument applies to the second series for $\nu<\nu_0$, so that the 
solution satisfying our boundary condition is 
$$
w_1(t,\nu) = \cases{\displaystyle
t^2\sum_{n=0}^\infty A^-_n(-1)^n{{(b)_n}\over{(c)_n}}
G_n(b-n,c,z)\left({\nu\over\nu_0}\right)^{-\alpha^-_n} & \qquad $\nu < \nu_0
\, ;$\cr
\, & \cr
\displaystyle t^2\sum_{n=0}^\infty A^+_n(-1)^n{{(b)_n}\over{(c)_n}}
G_n(b-n,c,z)\left({\nu\over\nu_0}\right)^{-\alpha^+_n} & \qquad $\nu \geq 
\nu_0\, .$\cr}\eqno(24)
$$

\beginfigure*{1}
\vskip 90mm \special{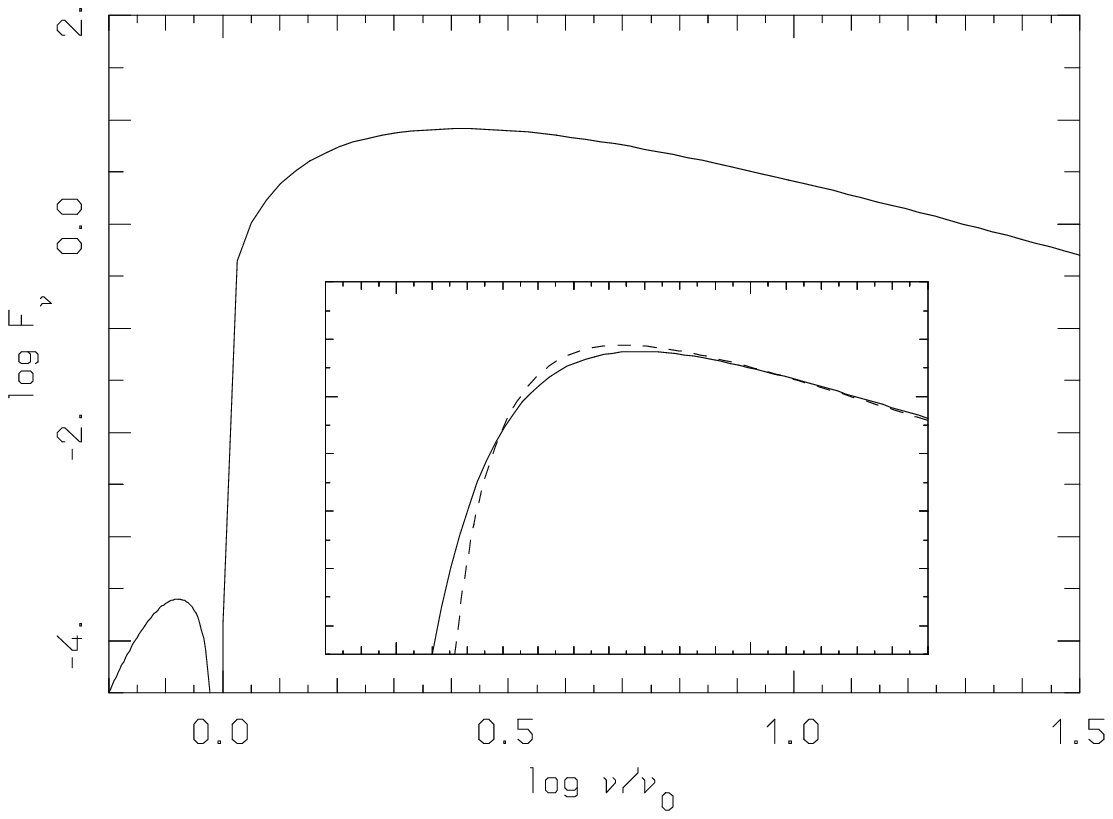}
\caption{{\bf Figure 1.}
Emergent flux $F_\nu$ (in arbitrary units) for spherical accretion onto 
a Schwarzschild black hole; here $t_h =20$, $t_*=0.9t_h$ and 
$\alpha_0^+ = 1.54$. 
A comparison with the analytical solution (dashed line)  
in the limit  $t_h \to \infty $ is illustrated in the box for 
$t_*=20$, showing a good agreement with PB result.}
\endfigure

Equation (24) exhibits two striking features, not shared by its
non--relativistic counterpart, which arise from the presence of 
advection/aberration terms in the moment equations. First of all, we
note that according to equation (24) photons injected at $\nu = \nu_0$ can 
be shifted {\it both\/} to higher and lower energies by dynamical 
Comptonization. This is in apparent contrast with PB result that photons can
only gain energy in scatterings with electrons in a converging flow
(the adiabatic compression).
PB statement is, however, correct up to $O(v)$ terms and their equation (8)
is the low--velocity limit of the more general expression for the photon energy
change along a geodesic (see e.g. Novikov \& Thorne 1973)
$${1\over \nu}\dert{\nu}{\ell} = -\left(n^i a_i + {1\over 3}\theta + n^in^j
\sigma_{ij}\right)\, ,\eqno(25)$$
where $n^i$ is the unit vector along the photon trajectory and $a^i$, $\theta$
and $\sigma_{ij}$ are the flow 4--acceleration, expansion and shear, 
respectively. In free--fall $a^i$ vanishes while it can be safely neglected 
in PB approximation being $O(v^2)$. The remaining two terms are both of
order $v$
$$\eqalign{ &\theta = -{3\over 2}{v\over r} \cr
              & n^in^j\sigma_{ij} =  {1\over 2}{v\over r}(3\mu^2 -1)\cr}
\eqno(26)
$$
where $\mu$ is the cosine of the angle between the photon and the radial 
directions. The mean photon energy change can be obtained 
angle--averaging equation (25) over the specific intensity
$$
I_\nu (\mu) = w_0 + 3 \mu w_1 + {15 \over 4 } (3\mu^2 - 1) w_2 + \ldots 
\, . \eqno(27)
$$        
Recalling the behaviour of the radiation moments in the 
diffusion limit, we get
$$\left < {1\over \nu}\dert{\nu}{\ell} \right >
=  {v \over r}\left[{1\over 2}  + {3\over\tau}\left({1\over\tau} - v\right)
\right]\, .\eqno(28)$$
The second term in square brackets arises because of shear and 
is negligible in PB approximation, being either $O(1/ \tau^2)$ or 
$O(v /\tau)$. This implies that the mean photon energy change is 
always positive. However, 
when advection and aberration are taken into 
account (see equations [25], [26]) photons moving in a cone around the 
radial direction suffer an energy loss and the collective effect is stronger  
when the flow velocity approaches unity 
in regions of moderate optical depth.  

The second important feature concerns the slope 
of the power--law, high--energy tail 
of the spectrum. From equation (22) the fundamental mode is 
$$
\alpha^+_0 = {-( 12 + 3 t_h) + \sqrt{ \left ( 12 + 3 t_h \right )^2 + 96 t_h }
\over 8 } \, , \eqno (29) 
$$
and, for large values of $t_h$, equation (29) gives
$$
\alpha^+_0 = 2 - {40 \over 3 t_h} 
+  {1120 \over 9 t_h^2} + O \left ( 1 / t_h^3 \right ) \, . \eqno (30) 
$$
At large enough frequencies the spectral index is dominated by 
the fundamental mode which, for $t_h\magcir 1$, 
sensibly deviates from the value predicted by the non--relativistic calculation.
Despite the fact that this effect is present below the trapping radius,
we stress that, contrary to a widespread belief, the trapping radius 
does not act as a one--way membrane. Photons produced near or below the
surface $\tau v =1$ can still escape to infinity
even if both the large optical depth and the strong advection caused by the 
inward flow dramatically reduce the emergent radiative flux. Moreover, these
photons, although comparatively few, are the more strongly 
comptonized and will anyway dominate the high--energy tail of the spectral
distribution. Equation (30) shows that
the emergent spectrum turns out to be flatter with respect to PB 
case. The two main features of our solution, harder spectrum and drift of 
photons below $\nu_0$, can be clearly seen in figure 1, where the emergent
spectrum is shown for $t_h=20$.

It is interesting to compare the present, analytical solution with the
numerical result obtained using the fully GR characteristic--ray code (CRM) 
described in Zane {\it et al.\/} (1996). 
In figure 2 we show the emergent spectrum relative
to the  ``cold'' solution for black hole accretion with $\rho_h = 1.42
\times 10^{-5}$ $\rm g\, cm^{-3}$. In this case both electron 
scattering and free--free emission/absorption are considered. At large enough 
frequencies scattering is the only source of opacity, so, in this limit, we 
expect our idealized analytical model to be representative of the realistic 
situation. The numerical model has $t_h\simeq 15$ which corresponds to 
$\alpha_0^+=1.43$. This value is in excellent agreement with the derived 
spectral index $\alpha = 1.36$.

\beginfigure*{2}
\vskip 90mm \special{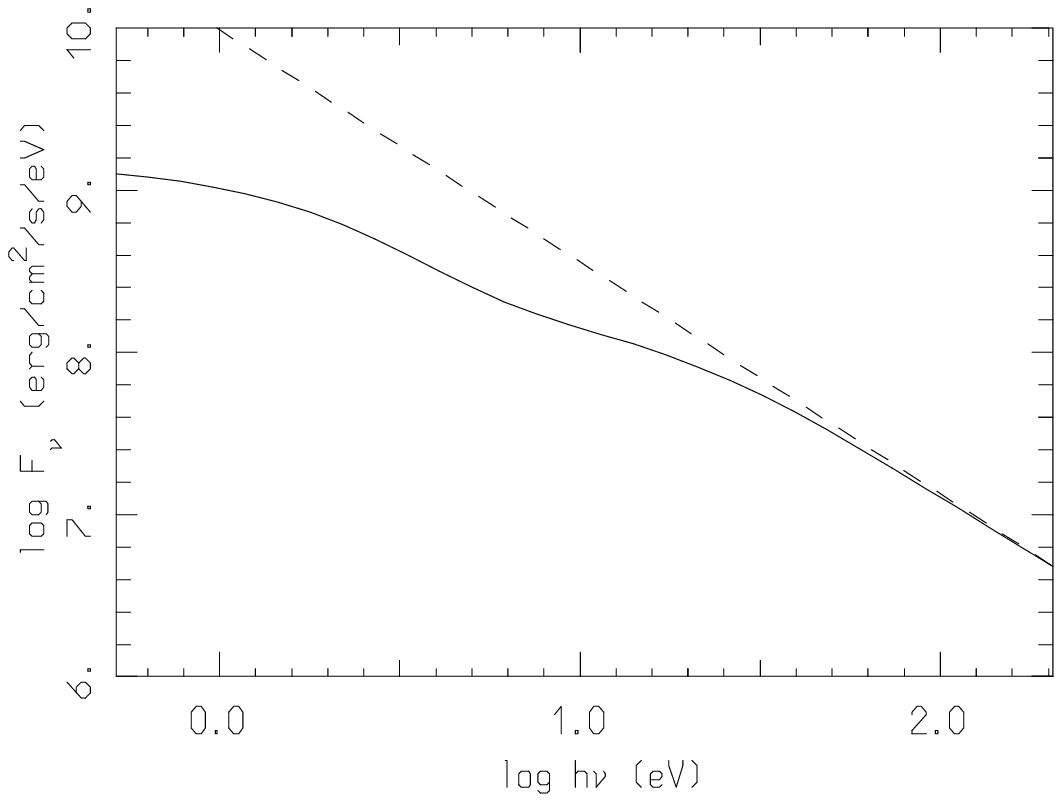}
\caption{{\bf Figure 2.}
Emergent flux $F_\nu$ computed using our CRM code; the derived spectral index 
is $1.36$. In this model $t_h\simeq 15$ and the corresponding value of
$\alpha_0^+$ is $1.43$ (dashed line).}
\endfigure

\section{Radiative transfer in an expanding atmosphere}

In this section we discuss 
the case of a pure scattering, expanding atmosphere with a power--law
velocity profile 
$$v = v_*\left({r\over {r_*}}\right)^\beta\eqno(31)$$
where the subscript ``$_*$'' refers to the base of the envelope, now 
$v$ is taken positive outwards and PB approximation is used. The 
second order partial differential equation for the radiation flux is 
given by equation (9), upon the substitution of $t$ with $-t$

$$t\derss{w_1}{t} +\left(t-1+\beta\right )\der{w_1}{t}-\left(1+{{2\beta
\over t}}\right)w_1 + {{2+\beta}\over 3}\der{w_1}{\ln\nu}=0\, .\eqno(32)$$

This equation could be integrated using the same technique discussed in 
section 2, looking for 
separable solutions 
$w_1 = t^2 h_1(t)\nu^{-\alpha}$. It can be easily
checked that equation (32) yields again, upon factorization, a Kummer equation
for $h_1(t)$, as in the converging flow case. 
A problem arises, however, as
far as boundary conditions are concerned: in section 2 
the only physically meaningful solution was selected asking that the 
flux become a constant for $t\to 0$ and that adiabatic 
compression of photons hold for $t\to\infty$. In that case the existence of
these physical constraints was sufficient to fix univocally the mathematical
solution. However, this particular issue turns out to be much 
more delicate in the wind problem. We preferred then to 
look for an alternative method of solution which allows for an 
easier handling of boundary conditions. Equation (32), describing
diffusion of photons through a moving medium, is a Fokker--Planck 
equation and can be brought into the standard Fokker--Planck
form 
$$\derss{(tu_1)}{t} -\der{[(-1-\beta-t)u_1]}{t}=\der{u_1}{x}
\, ,\eqno(33)$$ 
where we have defined $w_1= t^2 u_1$ and $x = -3/(2+\beta)\ln\nu$. 
The solution
can be found by Fourier transforming equation (33) with respect to $t$, 
solving the equation for the Fourier transform $\hat u_1$ and then
transforming back (see e.g. Risken 1989). The equation for the Fourier 
transform is obtained
from equation (33) replacing $\partial/\partial t$ by $ik$ and $t$ by
$i\partial/\partial k$,
$$ik(1+ik)\der{\ln\hat u_1}{ik} + \der{\ln\hat u_1}{x} = (1+\beta)ik\,
.\eqno(34)$$
This is a first order PDE which can be solved by standard methods
(see e.g. Sneddon 1957) once a boundary condition is specified. If we
assume that monochromatic photons of frequency $\nu_0$ are 
injected at $t=t _*$, the boundary condition for equation (34) is just
$\hat u_1 = u_1^0\exp (-ikt_*)$, here $u_1^0$ is the luminosity emitted
at $\nu_0$, and the corresponding solution is 
$$\hat u_1 = u_1^0\left\{\left[1-\exp(x - x_0)\right]ik+1\right\}^{1+\beta}
\exp\left\{-{{\exp(x - x_0)ikt_*}\over{\left[1-\exp(x - x_0)\right]ik+1}}
\right\}\, .\eqno(35)$$
The solution to equation (33) is given by the Fourier integral
$$u_1 = {1\over{2\pi}}\int^{+\infty}_{-\infty}\hat u_1\exp(ikt)\, dk$$
which can be evaluated analytically in terms of the modified Bessel function 
$I_q$ (see Appendix B)
$$\eqalign{ u_1 = & 
u_1^0\left({\nu\over{\nu_0}}\right)^{3/2}
\left({{t_*}\over t}\right)^{(2+\beta )/2}\left[1-\left({\nu\over{\nu_0}}
\right)^{3/(2+\beta )}\right]^{-1}\exp\left [-{{t+\left(
\nu/\nu_0\right)^{3/(2+\beta )}t_*}\over
{1-\left(\nu/\nu_0\right)^{3/(2+\beta )}}}\right ] \times \cr
& I_{2+\beta}\left[2\sqrt{(\nu/\nu_0)^{2/(3+\beta )}t_*t}\biggl /\left (1-
\left(\nu/\nu_0\right)^{2/(3+\beta )}\right )\right ]\, .
\cr}\eqno(36)$$
The main advantage of solving equation (32) following the method outlined here 
is that the Fourier transform is automatically selecting the regular solution,
because it can be computed only for functions that are $L_2$ in $]-\infty,
\infty[$. 
In other words, it is the method of solution itself which is suited for 
finding only regular solutions and in doing so no extra constraint is required.

The spectrum is shifted towards lower frequencies and it broadens at the same 
time, developing a power law, low--energy tail. The overall behaviour is 
similar to that of the converging flow but somehow reversed, since now 
photons can 
drift only to frequencies lower than $\nu_0$. There is, however, a major 
difference in the power--law index $\alpha$ between 
the two cases since $\alpha$ does not 
depend on $\beta$ for the wind solutions, as can be seen examining the 
spectral behaviour of equation (36) at low frequencies.
Since $I_q(z)\sim (z/2)^q/
\Gamma (q+1)$ when the argument is small,  we have for the emergent luminosity
$$L_\nu\propto \left({\nu\over {\nu_0}}\right)^3\exp\left[-{{
\left(\nu/\nu_0 \right )^{(2+\beta )/3}t_*}\over{1 - \left(\nu/\nu_0 \right )^
{(2+\beta )/3}}}\right]\left[1 - \left({\nu\over{\nu_0}} \right )^
{(2+\beta )/3}\right]^{-3-\beta}\sim \left({\nu\over{\nu_0}}\right)^3
\eqno(37)$$
if $\nu\ll\nu_0$, which shows that $\alpha=-3$ 
irrespective of the value of $\beta$. The monochromatic flux at $t=0$ is shown
in figure 3 for $\beta=1$ and $t_*=1$; the power--law tail at low energies is
clearly visible.

\beginfigure*{3}
\vskip 90mm \special{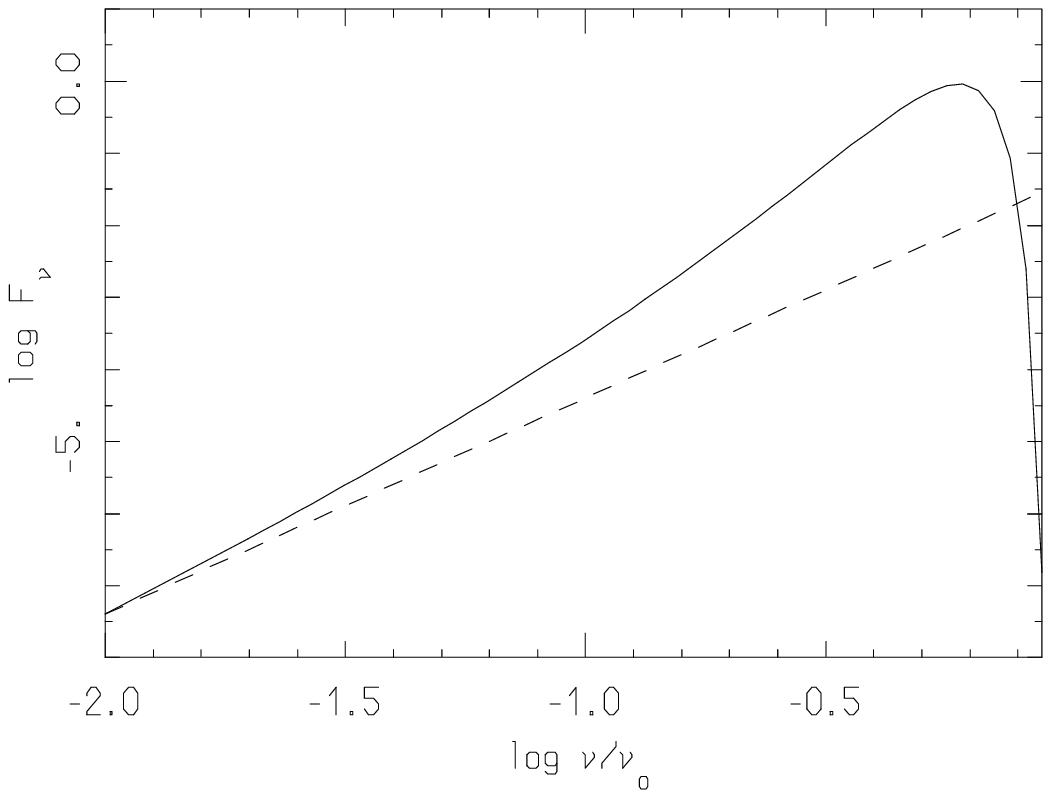}
\caption{{\bf Figure 3.}
Emergent flux $F_\nu$ (in arbitrary units) for an expanding envelope 
with $\beta =1$ and $t_*=1$.
At low energies the spectral index approaches $-3$ (dashed line).}
\endfigure

\section{Discussion and conclusions}

In this paper we have reconsidered the transfer of radiation in a
scattering, spherically--symmetric  medium, extending PB analysis of 
converging flows 
to the relativistic case and investigating the effects of bulk motion 
comptonization in expanding atmospheres. 
In the low--velocity limit and assuming diffusion approximation, PB found that
monochromatic photons injected at the base of the atmosphere always gain energy 
as they propagate outwards. The emergent spectrum exhibits an overall
shift to higher frequencies and a power--law, high--energy tail with a
spectral index depending on the velocity gradient.
Under the same assumptions, the wind solution shows similar, although reversed, 
features. Adiabatic expansion now produces an 
overall drift toward lower energies and the formation of 
a power--law tail at low frequencies. In this case, however, the spectral
index is independent on the velocity gradient and turns out to be always 
equal to $-3$.

Both these analyses are correct to first order in $v/c$ and can be thought to
adequately describe situations in which bulk motion is non--relativistic
in regions of moderate scattering depth. Relativistic corrections are, in fact,
related to the anisotropy of the radiation field and are washed out 
if $\tau \gg 1$. Obviously, if the flow is optically thin, repeated scatterings
are ineffective no matter how large the velocity is. In outflowing atmospheres
high velocities are expected at large radii, where the optical depth has 
dropped below unity, so that our assumption can be reasonable.
On the other hand, in accretion flows onto
compact objects the condition $\tau\gg 1$ where $v\sim 1$ is likely to 
be met only when the accretion 
rate becomes hypercritical. This shows that a relativistic treatment of
dynamical comptonization is indeed required in investigating the emission
properties of accretion flows. For $v\sim 1$ the diffusion limit is not
recovered simply asking that the radiative flux is proportional to the gradient
of the energy density, since the radiative shear is as important as the flux.
Relativistic corrections produces two main effects:
first, photons are shifted toward {\it both} higher and lower frequencies by 
dynamical comptonization and, second, the spectrum at large frequencies
is sensibly flatter than in the non--relativistic case. The spectral
index now depends not only on the velocity gradient, but also on the
value of the scattering depth at the horizon and goes 
to its non--relativistic limit when $\tau_h$ tends to infinity. 
Despite the fact that relativistic effects are important only where $\tau v 
> 1$, 
that is to say below the trapping radius, their signature is still present 
in the emergent spectrum. In particular, the high energy tail 
is populated by the strongly comptonized photons coming just from this region.
A similar effect was found by Mastichiadis \& Kylafis (1992, see also Zampieri,
Turolla, \& Treves 1993) in an accretion
flow onto a neutron star. In their case the formation of an essentially
flat ($\alpha\simeq 0$) spectrum is due to the fact that photons experience 
a very large number of energetic scatterings before emerging to infinity,
since no advection is present being the star surface a perfect reflector. 
Our spectrum is softer with respect to Mastichiadis \& Kylafis just because
a sizable fraction of the more boosted photons are dragged into the hole,
but, at the same time, it is harder than PB since 
in the relativistic regime the mean energy gain per scattering is higher.
From the mathematical point of view it is noteworthy that the assumption of 
a finite optical depth at the inner boundary 
(i.e. at the horizon in our model or at the reflecting surface in Mastichiadis 
\& Kylafis) produces a fundamental mode which is flatter 
with respect to PB; in both cases PB result is recovered in the limit 
$\tau_h \to \infty$. 
The possibility that scattering of photons in an accretion flow onto
a black hole produces a power--law tail with spectral index flatter than 2
was also suggested in a very recent paper by Ebisawa, Titarchuk, \& Chakrabarti
(1996). Using a semi--qualitative analisys they found that the spectral
index is close to 3/2 for large values of the optical depth at the horizon
and discussed the possible relevance of this result in connection with the
observed hard X--ray emission from black hole candidates in the high state.
 
We note that for $1<\tau_h<32/9$ the predicted spectral index (see equation
[29]) is smaller than $1$, implying a divergent frequency--integrated 
luminosity; this behaviour is not new and was already found by Schinder \& 
Bogdan (1989) and Mastichiadis \& Kylafis (1992). It simply reflects
the fact that photons can gain an {\it arbitrarily large\/} amount of
energy in collisions with the free--falling electrons.
It should be taken into account, however, 
that, when $h\nu\approx m_ec^2$ the electron recoil in the particle
rest frame can not be neglected anymore, so for large enough energies our
treatment is not valid, as discussed in more detail in Zampieri (1995).
The decrease of the cross--section in the quantum
limit makes the scattering process less efficient, producing a sharp cut--off
in the spectral distribution.
Finally, as already stressed by Blandford \& Payne (1981a) and Colpi (1988),
thermal comptonization dominates over dynamical comptonization when 
$v^2\mincir 12kT/m_e$. The spectral distribution depends then
on the relative strength of competitive processes such as 
heating/cooling by thermal comptonization and compressional heating and must 
be derived solving the radiative transfer equation in its complete form.  

\section* {References}

\beginrefs

\ref{Abramowitz, M., \& Stegun, I.A. 1972, Handbook of Mathematical
Functions, (New York: Dover), AS}
\ref{Blandford, R.D., \& Payne, D.G. 1981a, MNRAS, 194, 1033}
\ref{Blandford, R.D., \& Payne, D.G. 1981b, MNRAS, 194, 1041}
\ref{Colpi, M. 1988, ApJ, 326, 233}
\ref{Cowsik, R., \& Lee, M.A. 1982, Proc. Roy. Soc. London A, 383, 409}
\ref{Ebisawa, K., Titarchuk, L., \& Chakrabarti, S.K. 1996, PASJ, 48, 59}
\ref{Gilden, D.L., \& Wheeler, J.C. 1980, ApJ, 239, 705} 
\ref{Mastichiadis, A., \& Kylafis, N.D. 1992, ApJ, 384, 136}
\ref{Nobili, L., Turolla, R., \& Zampieri, L. 1991, ApJ, 383, 250}
\ref{Nobili, L., Turolla, R., \& Zampieri, L. 1993, ApJ, 404, 686}
\ref{Novikov, I.D., \& Thorne, K.S. 1973, in Black Holes, DeWitt, C. \&
DeWitt B.S. eds., (New York: Gordon \& Breach)}
\ref{Payne, D.G., \& Blandford, R.D. 1981, MNRAS, 196, 781, PB}
\ref{Prudnikov, A.P., Brychkov, Yu.A., \& Marichev, O.I. 1986, Integrals
and Series, Vol. I (New York: Gordon \& Breach)}
\ref{Risken, H. 1989, The Fokker--Planck Equation (Berlin: Springer--Verlag)}
\ref{Schneider, P., \& Bogdan, T.J. 1989, ApJ, 347, 496}
\ref{Sneddon, I.N. 1956, Special Functions of Mathematical Physics and
Chemistry (Edinburgh: Oliver \& Boyd)}
\ref{Sneddon, I.N. 1957, Elements of Partial Differential Equations (New
York: McGraw--Hill)}
\ref{Thorne, K.S. 1981, MNRAS, 194, 439}
\ref{Thorne, K.S., Flammang, R.A., \& \.Zytkov, A.N. 1981, MNRAS, 194, 475}
\ref{Turolla, R., \&  Nobili, L. 1988, MNRAS, 235, 1273}
\ref{Zampieri, L., Turolla, R., \& Treves, A. 1993, ApJ, 419, 311}
\ref{Zampieri, L. 1995, unpublished PhD Thesis}
\ref{Zampieri, L., Miller, J.C., \& Turolla, R. 1996, MNRAS, in the press}
\ref{Zane, S., Turolla, R., Nobili, L., \& Erna, M. 1996, ApJ, in the press}

\endrefs

\section*{Appendix A}


Since the polynomials $G_n (b-n, c, z)$ appearing in equation (24) are not 
an orthogonal system, is not possible to derive an explicit expression 
for the coefficients $A^\pm_n$. Here we show that these constants can be, 
in principle, obtained as the solution of an infinite system of 
linear algebraic equations. We note that the two sets $A^\pm_n$ are not 
independent, because the two expressions in (24) must match 
at $\nu = \nu_0$, where 
$$
w_1(t, \nu_0) = A \delta ( t/t_* - 1)\eqno({\rm A}1)
$$
($A$ is a constant related to the monochromatic flux injected at the inner 
boundary). Since $z\propto t/t_h$ and $x=t/t_*$, the polynomials $G_n(b-n,c,z)$ 
can be expressed in terms of $G_n(3,3,x)$, which form an orthogonal system, as
$$
G_n(b-n,c,z) = \sum_{m=0}^n C_{nm}G_m(3,3,x)\, .\eqno({\rm A}2)
$$
The coefficients $C_{nm}$ are solution of the upper triangular system of 
linear algebraic equations
$$
\sum_{m=k}^n (-1)^{m-n}{m\choose k}{{(m+2)!(m+2+k)!}\over{(2m+2)!
(k+2)!}}C_{nm} = {n\choose k}{{\Gamma(c+n)\Gamma(b+k)}\over{\Gamma(b+n)
\Gamma(c+k)}}\left(-{\beta\over\gamma}t_*\right)^k\qquad\qquad k=0,\ldots , n\, .
\eqno({\rm A}3)
$$
Recalling the standard expansion of the $\delta$--function over an orthogonal
set of eigenfunctions and using again $G_n(3,3,x)$ as a basis, it is
$$
\delta (x-1) = x^2\sum_{m=0}^\infty {{(2m+3)!}\over{m!(m+2)!}}G_m(3,3,x)\, .
\eqno({\rm A}4)
$$
Inserting (A2) and (A4) into (A1) and equating the coefficients of the
polynomials of the same degree, we obtain
$$
\sum_{n=m}^\infty (-1)^n {{(b)_n}\over{(c)_n}}C_{nm}A^\pm_n =
{A\over{t_*^2}}{{(2m+3)!}\over{m!(m+2)!}}\qquad\qquad m\geq 0\, .
\eqno({\rm A}5)
$$
The numerical evaluation of $A^\pm_n$ has been carried out truncating the
series appearing in (A5) to a maximum order $N\sim 60$ and solving the
system by backsubstitution.

\section*{Appendix B}


In this Appendix we derive the expression for $u_1$, equation (36), starting 
from the Fourier integral. By defining, for the sake of conciseness,
$a = (\nu/\nu_0)^{3/(2+\beta )}$, the Fourier integral can be written as
$$u_1 = {u_1^0\over{2\pi}}\int_{-\infty}^{+\infty}\left[\left(1-a\right)ik+1
\right]^{1+\beta}\exp\left[ikt - {{aikt_*}\over{(1-a)ik+1}}\right]\, dk\, ,
\eqno(\rm B1)$$
which can be transformed into an integral in the complex plane by introducing
the new, complex, integration variable $z = (1-a)ik+1$:
$$u_1 = {u_1^0\over{2\pi i}}{1\over{1-a}}\exp\left[-{{t+at_*}\over{1-a}}\right]
\int_{-i\infty+1}^{+i\infty+1}z^{1+\beta}
\exp\left[{t\over{1-a}}z + {{at_*}\over{1-a}}z^{-1}\right]\, dz\, .
\eqno(\rm B2)$$
The integral appearing in (B2) defines the Bessel function of imaginary
argument (see Prudnikov, Brychkov, \& Marichev 1986), 
$$\int^{+i\infty +1}_{-i\infty +1}z^{1+\beta}
\exp\left[{t\over{1-a}}z + {{at_*}\over{1-a}}z^{-1}\right]\, dz\, = 
2\pi i\left({{at_*}\over t}\right)^{(2+\beta)/2} 
J_{2+\beta}[i2(att_*)^{1/2}/(1-a)] \eqno(\rm B3)$$
so that finally we have
$$u_1 = {{u_1^0}\over{1-a}}\left({{at_*}\over t}\right)^{(2+\beta)/2}
\exp\left[-{{t+at_*}\over{1-a}}\right]
I_{2+\beta}\left[2{{(att_*)^{1/2}}\over{1-a}}\right]\, ,\eqno(\rm B4)$$
which is exactly equation (36). 
We note that, although the absolute 
convergence in the complex plane of the integral representation
(B3) is proved only for $at_*/(a-1)>0$, $2+\beta < 1$, 
direct substitution of (B4) into the Fokker--Planck equation
(33) shows that (B4) is a solution with the only restriction 
$a < 1 $.

\bye